\documentclass[12pt]{article}
\usepackage{latexsym}
\usepackage{epsfig,amssymb,euscript}
\usepackage{amsmath}

\def\be{\begin{equation}}
\def\ee{\end{equation}}
\def\bea{\begin{eqnarray}}
\def\eea{\end{eqnarray}}

\input epsf
\begin{document}
\font\cmss=cmss10 \font\cmsss=cmss10 at 7pt
\hfill SISSA-30/2005/EP
\par\hfill LPTHE-05-11
\par\hfill UB-ECM-PF-05/14
\par\hfill hep-th/0505055
\\
\vspace{15pt}
\begin{center}
{\LARGE \bf Supersymmetry breaking} 
\vskip 25pt
{\LARGE \bf at the end of a cascade of Seiberg dualities} 
\end{center}

\vspace{10pt}

\begin{center}

{\Large M. Bertolini$\,^a$, F. Bigazzi$\,^{b,c}$, A. L. Cotrone$\,^{d}$} \\ 
\end{center}
\vskip 10pt
\begin{center}
\textit{a} SISSA/ISAS and INFN, Via Beirut 2; I-34014 Trieste, Italy.\\
\textit{b} LPTHE, Universit\'es Paris VI and VII, 4 place Jussieu; 75005, Paris, France.\\
\textit{c}  INFN, Piazza dei Caprettari, 70; I-00186  Roma, Italy.\\
\textit{d} Departament ECM, Facultat de F\'isica, Universitat de Barcelona and \\Institut 
de Fisica d'Altes Energies, Diagonal 647, E-08028 Barcelona, Spain.\\

{\small\tt bertmat@sissa.it, bigazzi@lpthe.jussieu.fr, cotrone@ecm.ub.es}
\end{center}

\vspace{15pt}

\begin{center}
\textbf{Abstract}
\end{center}

\vspace{4pt} 
{\small \noindent
We study the IR dynamics of the cascading non-conformal quiver theory on $N$ regular and $M$ fractional D3 branes 
at the tip of the complex cone over the first del Pezzo surface. The horizon of this cone is the irregular 
Sasaki-Einstein manifold $Y^{2,1}$. Our analysis shows that at the end of the cascade supersymmetry is 
dynamically broken.}
\vfill

\newpage

\section{Introduction and main results}
Recently, a new class of $AdS/$CFT dual pairs
has been found \cite{gmswSE,MS,MS2} and a number of duality checks have been successfully performed \cite{bbc,MS2}. 
The ten-dimensional type IIB supergravity background has $AdS_5 \times Y^{p,q}$ geometry, constant $F_5$-flux 
through $Y^{p,q}$ and constant dilaton and axion. $Y^{p,q}$ are five-dimensional Sasaki-Einstein manifolds, 
where $q<p$ are positive coprime integers. 

The dual gauge theories are four dimensional ${\cal N}=1$ SCFT's describing the low energy dynamics of a 
bunch of $N$ D3-branes placed at 
the tip of the Calabi-Yau cone $C$ over $Y^{p,q}$. 
The moduli space of the theory is described by $N$ copies of the defining equations of the cone.
Since the topology of the $Y^{p,q}$ manifolds is $S^3\times S^2$ 
one can consider the possibility of adding fractional branes to this system. The latter can be thought of 
as D5-branes wrapped on $S^2$. From the gauge theory point of view this breaks conformal 
invariance. From the supergravity point of view this changes the geometry drastically. 
In particular, the geometry of the cone is expected to be deformed since in general a deformation should 
occur in the field theory moduli space. Moreover,
the complex type IIB three-form $G_{3} = F_{3} + \tau H_{3}$ acquires a non-trivial profile and the RR 
five-form flux a radial dependence.

A first crucial step towards finding the complete supergravity solution corresponding to a bunch of $N$ regular 
and $M$ fractional D3-branes has been taken in \cite{HEK}.
As a simplifying assumption, the geometry of the cone was taken unchanged with respect to the conformal case.
The results in \cite{HEK} match with the dual field theory expectations in the UV, in the limit $M<<N$. Still, the solution 
has a singularity in the region corresponding to the IR 
of the dual field theory. Showing the existence of a smooth deformation 
of the geometry remains an open problem. Actually, there are geometrical arguments against the existence 
of complex deformations of the cone $C(Y^{p,q})$ preserving supersymmetry \cite{FHU}, though in \cite{leo} 
a first order approximation to such a deformation has been found.

The structure of the solution found in \cite{HEK}, in particular the fact that the $F_5$-flux is not constant, 
suggests that duality cascade phenomena might take place as the theory flows through the IR, in analogy to the 
conifold case \cite{KT,KS}. For the cases $Y^{p,p-1}$ and $Y^{p,1}$, 
the cascade can reach a point where the gauge group becomes $SU(M) \times SU(2M) \times \dots \times SU(2pM)$ and 
for the $SU(2pM)$ factor there are effectively $2pM$ flavors. 
This implies that the moduli space of the theory will receive quantum corrections and so a deformation of the 
Calabi-Yau cone should occur. 

In this paper we try to understand this deformation from a dual perspective. We analyze, from a pure field theory 
point of 
view, the case of the theory for $Y^{2,1}$ (in this case the Calabi-Yau is the complex cone over the first del 
Pezzo surface $dP_1$), which can be seen as a master example for both $Y^{p,p-1}$ and $Y^{p,1}$ cases. 
Our results differ sensibly from what happens for the conifold. Most notably, the theory does not have a 
supersymmetric vacuum. 
Therefore, a dual smooth supergravity background, 
if it exists, should correspond to a non-supersymmetric deformation of the singular geometry. 
This result, which is likely to hold for the full $Y^{p,p-1}$ and $Y^{p,1}$
series, opens up the very interesting possibility of dealing with an all new
class of non-$AdS$/non-CFT dual pairs with broken supersymmetry, where a
number of non-supersymmetric field theory IR phenomena could be tackled in
a well defined setting.

This note is organized as follows. In section 2 we briefly review the structure of the gauge theory 
for $dP_1$ recalling how the cascade of Seiberg dualities takes place. In section 3, which contains 
our main results, we follow the cascade step by step up to the point where quantum deformations of the 
moduli space are expected. We thus study in detail this point and show that supersymmetry is 
dynamically broken. 

\vskip 10pt
{\bf Note added}: While this paper was being completed, two works appeared \cite{BHOP,FHSU} which 
address similar issues. Their conclusions agree with ours.

\section{Regular and fractional D3-branes on $\mathbf{C(Y^{2,1})}$}
Let us place $N$ D3-branes at the tip of $C(Y^{2,1})$, which is the complex 
cone over the first del Pezzo surface $dP_1$ \cite{MS}. The conformal field theory has gauge 
group $SU(N)^4$, bi-fundamental matter 
and a marginal superpotential \cite{hanany}. The $AdS/$CFT correspondence was checked for this case in \cite{bbc} 
where the exact R-charges and the central charge of the theory were computed using a-maximization and shown to agree 
exactly with the dual supergravity predictions made in \cite{MS}. 

The addition of $M$ fractional D3-branes (i.e. D5-branes wrapped on $S^2$) was considered in \cite{FHHW}. 
The corresponding quiver diagram is reported in figure 1. 
\begin{figure}[ht]
\begin{center}
{\includegraphics{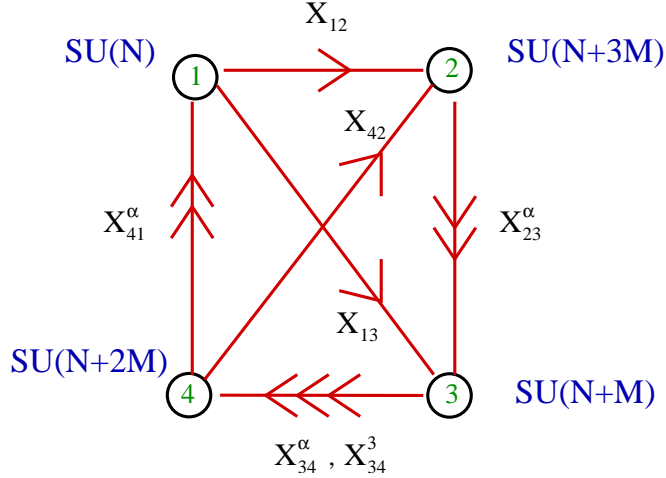}}
\caption{\small The non conformal quiver associated to the first del Pezzo 
surface. Each node represents a gauge factor. Each arrow represents a bi-fundamental chiral multiplet.  
The chiral fields $X_{41}^\alpha~,~X_{34}^\alpha~,~X_{23}^\alpha$ are doublets with respect to the $SU(2)$ 
flavor symmetry. 
The quiver for the conformal case is just the same, with $M=0$.}
\label{reg1}
\end{center}
\end{figure}
There is also a superpotential 
\be
\label{super}
W= \mbox{Tr}\;\left[\epsilon_{\alpha\beta}\,X^\alpha_{34} X^\beta_{41}X_{13}-\epsilon_{\alpha\beta}\, 
X^\alpha_{34} X_{42}X^\beta_{23} + \epsilon_{\alpha\beta}\,X^3_{34} X^\alpha_{41}X_{12}X^\beta_{23}\right],
\ee
which breaks the global symmetry group to $SU(2) \times U(1) \times U(1)$. The chiral fields 
$X_{41}^\alpha~,~X_{34}^\alpha~,~X_{23}^\alpha$ are doublets with respect to the $SU(2)$ symmetry

Let us consider some of the properties of this non conformal theory, beginning with the $\beta$-functions. We will 
take $M<<N$ in the following. For any node the $\beta$-function is proportional to
\be
b \equiv [3T(G)-\sum_iT(r_i)(1-\gamma_i)],
\ee
where $\gamma_i$ are the conformal dimensions of the fields and $T(G)$, $T(r_i)$ 
are the Casimir of the adjoint and $r_i$ representations.

The anomalous dimensions $\gamma_i$ of the bi-fundamental fields in the theory
are expected to differ from the ones at the conformal fixed point ($M=0$ case)
by factors of order $(M/N)^2$. This is because the quiver in figure 1 is invariant under 
$N \rightarrow N+3M, M \rightarrow -M$. In the limit $M<<N$ this corresponds to 
$N \rightarrow N, M \rightarrow -M$, hence the anomalous dimensions cannot depend linearly on $M/N$. 
This means that in order to give the $\beta$-functions for each node of the quiver at leading order in $M/N$, 
we will only need to know the anomalous dimensions (or the exact R-charges $3R_i=\gamma_i+2$) for 
the conformal case. 

Using the fact that the $\beta$-functions are zero in the conformal case, we end up with the following results 
\bea
\label{beta}
b_1 &=& {3M\over2}[4(R_{14}-1)+(R_{13}-1)+3(R_{12}-1)],\nonumber \\
b_2 &=& 3M [3+ (R_{23}-1)+(R_{24}-1)],\nonumber \\
b_3 &=& 3M [1+ 3(R_{23}-1)+ 2(R^{(\alpha)}_{34}-1)+(R^{(3)}_{34}-1)],\nonumber \\
b_4 &=& {3M\over2} [4+ 3(R_{24}-1)+ 2(R^{(\alpha)}_{34}-1) +(R^{(3)}_{34}-1)],
\eea
where $R_{ij}$ are the exact R-charges of the bi-fundamentals $X_{ij}$ in the $M=0$ case \cite{bbc}
\begin{eqnarray}
R(X_{12})&=& \frac 13 (-17 + 5 \sqrt{13}),\quad   
R(X^\alpha_{23})= R(X^\alpha_{41})=\frac 43 (4 - \sqrt{13})\nonumber \\  
R(X^\alpha_{34})&=& \frac 13 (-1 + \sqrt{13}), \quad   
R(X^3_{34})= R(X_{13})=R(X_{42})= - 3 + \sqrt{13} 
\end{eqnarray}

Thus we find
\bea
b_1 &=& -M(10-\sqrt{13}),\qquad \,\,\, b_2=-b_1 >0,\nonumber \\
b_3 &=& -M(7\sqrt{13}-22), \qquad b_4=-b_3 > 0.
\eea
There are couples of couplings running in opposite directions. Node (2) (whose gauge factor is $SU(N+3M)$) 
will generically run to 
infinite coupling first. To follow the theory at smaller energy scales one can Seiberg dualize on this node, and 
proceed. Actually there will be a cascade of Seiberg dualities in which the number of colors will get smaller and 
smaller values, but the structure of the quiver and the superpotential will remain unchanged
(until some node has $N_f = N_c$). A throughout analysis of how this occurs will be made in section 
\ref{selfsim}. 

All we have discussed in this section can be generalized to the cases $Y^{p,p-1}$ and $Y^{p,1}$, for which a similar dynamics takes place.

\section{The duality cascade}
In the cases $Y^{p,p-1}$, $Y^{p,1}$ the cascade of Seiberg dualities evolves to a theory with gauge 
group $SU(M)\times SU(2M)\times....\times SU(2pM)$ \cite{HEK}. The last node has $N_f=N_c=2pM$ and hence a 
non-perturbative 
modification of the moduli space of the theory will occur. Before 
this point is reached all the nodes have $N_f>N_c$ and the moduli space should not differ from the original one, 
that is (copies of) the cone over $Y^{p,q}$.

A counting of the number of flavors in the $Y^{2,1}$ case is useful at this point. If we consider the original 
theory, we find that the nodes of the $SU(N)\times SU(N+3M)\times SU(N+M)\times SU(N+2M)$ quiver have the 
following number of fundamental and anti-fundamental fields, respectively
\be
N^{(1)}_f= 4M+2N,\quad N^{(2)}_f= 2M+2N,\quad N^{(3)}_f= 3N+6M,\quad N^{(4)}_f= 3N+3M. 
\ee
Thus, starting with $N$ being a multiple of $M$, as it decreases and eventually reaches $N=M$, we have that 
node $(2)$ ends up with gauge group $SU(4M)$ and $N_f=N_c=4M$. Hence one expects the moduli space to be modified.

Ignoring this fact, the last step  
of the cascade would reduce the number of gauge groups to three, giving a $SU(M)\times SU(2M)\times SU(3M)$ theory 
(i.e. node (1) has disappeared, see figure 1). The gauge group $SU(3M)$ would have $N_f=2M<N_c$ and thus the 
related gauge theory (if one forgets the superpotential induced by the previous step) would not be expected to 
have a supersymmetric vacuum. As we will show in the following, a careful analysis of the cascade does not 
change this conclusion, though the physical picture is slightly different.

\subsection{The self-similarity of the superpotential}\label{selfsim}

It is known that the complex cone over $dP_1$ has only one toric phase \cite{hanany}. 
This corresponds to the fact that 
(modulo rotations of the indexes) there is only one conformal field theory 
model corresponding to $dP_1$. This means that the theory on $dP_1$ is self-similar, and this property is expected to 
be preserved in the non-conformal case as far as the Seiberg dual theory does not have nodes with $N_f$ less or equal 
to $N_c$. Hence, at each step of the cascade, the quiver 
diagram looks the same and similarly the superpotential, up to the value of the superpotential coupling $\lambda$.
For completeness, we will now prove the latter fact explicitly.

Let us consider first the conformal case $M=0$. The superpotential is formally the same as in 
(\ref{super}).\footnote{We will
henceforth call $X_{ij} \equiv X_{ij}^1, Y_{ij} \equiv X_{ij}^2, Z_{ij} \equiv
X_{ij}^3$} Let us consider the effect of a Seiberg duality on node $(2)$. 
Since $N_f^{(2)}=2N$ the duality gives a 
new node $(2)$ with $N_c=N$, as before, and equal number of flavors. The new node $(2)$ will have anti-fundamental fields 
  $x_{32}, y_{32}$ and 
fundamentals $x_{21}, x_{24}$. Moreover there will be a meson matrix which, in terms of the old variables will have block 
components $M^X_{13}=X_{12}X_{23}$,  $M^Y_{13}=X_{12}Y_{23}$, $M^X_{43}=X_{42}X_{23}$, $M^Y_{43}=X_{42}Y_{23}$.

The superpotential inherited from the original theory will read
\bea
W &=& X_{34}Y_{41}X_{13} -Y_{34}X_{41}X_{13}- X_{34}M^Y_{43}+ Y_{34}M^X_{43}+\nonumber \\
&& Z_{34}X_{41}M^Y_{13}- Z_{34}Y_{41}M^X_{13}.
\label{W0b}
\eea
Moreover there will be a superpotential term (we redefine the fields so that the dimensional parameter $1/\mu$ 
is re-absorbed; notice that we have also re-absorbed the $W$ coupling above)
\bea\label{extra}
Tr M q {\tilde q} =  M^X_{13}\,x_{32}\,x_{21}- M^Y_{13}\,y_{32}\,x_{21} + M^Y_{43}\,y_{32}\,x_{24}- 
M^X_{43}\,x_{32}\,x_{24}.
\eea
The choice on the signs is such that they are the opposite of the terms in $W$ containing the same mesons.
Of course, they preserve the $SU(2)$ flavor symmetry of the theory. 

The total superpotential for the dual theory thus reads
\bea
W_{tot}&=& X_{34}( Y_{41}X_{13} - M^Y_{43}) -Y_{34}( X_{41}X_{13} - M^X_{43}) +M^Y_{13}(Z_{34}X_{41} - y_{32}\,x_{21})\nonumber \\
&&  -  M^X_{13}(Z_{34}Y_{41} - x_{32}\,x_{21}) +  M^Y_{43}\,y_{32}\,x_{24} - M^X_{43}\,x_{32}\,x_{24}.
\eea
By using the F-term equations w.r.t. $X_{34}, Y_{34}$, $M^X_{43}$, $M^Y_{43}$ we see that these fields are all 
massive and can be integrated out. These F-term equations give
\be
Y_{41}\,X_{13} = M^Y_{43} , \quad  X_{41}\,X_{13}= M^X_{43} , \quad y_{32}\,x_{24} = X_{34} , \quad x_{32}\,x_{24} = Y_{34} ,
\ee
which substituted in the above superpotential give
\bea
W_{tot} &=& M^Y_{13}(Z_{34}X_{41}-y_{32}\,x_{21}) - M^X_{13} (Z_{34} Y_{41} -x_{32}\,x_{21}) +\nonumber \\
&& Y_{41}X_{13}\,y_{32}\,x_{24} - X_{41}X_{13}\,x_{32}\,x_{24}.
\eea
This superpotential is exactly equivalent to the original one $W$, Eq.(\ref{super}). In fact, the whole 
theory is equivalent 
to the original one. This is evident from the following redefinition of fields
\bea
&& M^Y_{13}={\hat Y}_{13}, \quad Y_{41} = {\hat Y}_{41}, \quad y_{32}={\hat X}_{32}, \quad X_{13}= {\hat Z}_{13}, 
\quad Z_{34} = {\hat X}_{34},\nonumber \\
&& M^X_{13}={\hat X}_{13}, \quad X_{41} = {\hat X}_{41}, \quad x_{32}={\hat Y}_{32}, \quad x_{21}={\hat X}_{21}, 
\quad x_{24}= {\hat X}_{24},
\eea
followed by the rotation of indexes
\be
(1)\rightarrow ({\hat 3}), \quad (2)\rightarrow ({\hat 1}), \quad (3)\rightarrow ({\hat 4}), \quad (4)\rightarrow ({\hat 2}),
\label{rede}
\ee
which give back simply the original theory.

In the non-conformal case, if we perform a Seiberg duality on node
$(2)$ there is no difference with respect to the  calculations above,
apart from the fact that now the gauge group on node $(2)$ goes from
$SU(N+3M)$ to $SU(N-M)$. Again, the  dual theory will be exactly as
the original one but with the shift $N\rightarrow N-M$. Performing a
second Seiberg duality on node $({\hat 2})$ we will find another
equivalent theory with $N\rightarrow N-2M$ with  respect to the
original one.\footnote{Remember that node $({\hat 2})$ is again the
one whose coupling diverges before the other ones', so the duality
must be taken on this node.}  After $k$ steps of the cascade
$N\rightarrow N-kM$. For every step of  the Seiberg cascade (before
the ``critical'' one where some node has $N_f$ equal or less than
$N_c$), we can take $W$ as in the original model, the only
differences, step by step, being in the superpotential coupling
$\lambda$. Hence we can treat each step of the cascade in a similar manner, 
as long as $N_f > N_c$ for every node.
If $N=lM$, after $l-1$ steps $N\rightarrow M$ and the field theory requires a more careful analysis.

\subsection{The cascade at $\mathbf{N=M}$ and beyond}\label{next}

Let us thus consider the cascade at
$N=M$, where we end up with a $SU(M)\times SU(4M)\times SU(2M)\times SU(3M)$ theory. 
Since the gauge coupling for 
the second node generically blows up before the others, it makes sense to consider an energy scale where 
the other nodes are weakly coupled.

The $4M$ fundamental and anti-fundamental flavors of the $SU(4M)$ theory are given by ($A=1,2,...4M$ 
is a color index, $a,k,i$ are flavor indexes)
\bea
Q &=& \left(~(X_{23})^A_a ~,~ (Y_{23})^A_a~\right), \quad a= 1,...,2M,\\
{\tilde Q}&=& \left(~(X_{12})^i_A~,~(X_{42})^k_A~\right), \quad k=1,2,...,3M, \quad i=1,...,M,
\eea
arranged as $4M\times 4M$ matrices.

Let us now use gauge and flavor invariance (which is
$SU(2)\times SU(2M)\times SU(3M)\times SU(M)$) so that, being painfully explicit with indexes, we get
\bea
Q &=&{\rm diag}\left(~(X_{23})^{1}_1,..., (X_{23})^{2M}_{2M},(Y_{23})^{2M+1}_1,...,(Y_{23})^{4M}_{2M}~\right), 
\nonumber \\
{\tilde Q} &=& {\rm diag}\left(~(X_{12})^1_1,...,(X_{12})^M_M, (X_{42})^1_{M+1},...,(X_{42})^{3M}_{4M}~\right).
\eea
The meson matrix ${\cal M}= Q\,\tilde Q$ will be given by
\bea
{\cal M}&=& {\rm diag}(~(X_{23})^1_1~(X_{12})^1_1~,~...~,~(X_{23})^M_M~(X_{12})^M_M~,~ 
(X_{23})^{M+1}_{M+1}~(X_{42})^1_{M+1}~,~...~,~\nonumber \\ &&(X_{23})^{2M}_{2M}~(X_{42})^M_{2M}~,~
(Y_{23})^{2M+1}_1~(X_{42})^{M+1}_{2M+1}~,~...~,~(Y_{23})^{4M}_{2M}~(X_{42})^{3M}_{4M}~).
\eea
We will use the following straightforward notations for the meson components\footnote{Notice that we can formally write
$\det {\cal M} = x^M\,y^{3M}$
after having identified $M^X_{13}$ with a complex parameter $x$, and $M^X_{34}, M^Y_{34}$ with $y$. 
The equation $\det {\cal M}=0$ thus seems to be related (after recovering the explicit $SU(2)$ invariance which is 
broken by the above parameterization) to $M$ copies of the complex cone over $dP_1$ which is in 
fact described by an equation of the form $x\,y^3=z\, w^3$ \cite{MS}.}
\be
M^X_{13}=X_{23}X_{12},\quad M^X_{34}=X_{23}X_{42},\quad  M^Y_{13}=Y_{23}X_{12}, \quad  M^Y_{34}=Y_{23}X_{42}.
\ee
The baryonic fields are formally given by
\be
B\approx (X_{23})^{2M}(Y_{23})^{2M}, \qquad {\tilde B}\approx (X_{12})^{M}(X_{42})^{3M}.
\ee
As it is well known, the theory without any other superpotential would have a moduli space described by 
$\det{\cal M}-B{\tilde B}=\Lambda^{2N_c}$.
In fact the 
theory has also a superpotential term inherited by the original model on $dP_1$, so that one has to consider 
the whole term  
\bea
W&=&\mbox{Tr}\;[X_{34} Y_{41}X_{13} - Y_{34} X_{41}X_{13}- X_{34} M^Y_{34}+ 
\nonumber \\&+& 
Y_{34}M^X_{34} + Z_{34} X_{41}M^Y_{13}-Z_{34} Y_{41}M^X_{13}]+\nonumber \\
&+&  \xi (\det{\cal M}-B{\tilde B}-\Lambda^{8M}), \label{totalW}
\eea
where $\xi$ is a Lagrange multiplier. 

Let us now fix gauge and global invariance on the other nodes 
of the quiver, i.e. solve the other three D-term equations.
\begin{itemize}
\item Node $(3)$ has $N_f=9M$ and $N_c=2M$. The fundamental flavors can be represented as a (symbolic) row 
$\left(X_{34},Y_{34},Z_{34}\right)$. Each block behaves as $n_f=3M$ flavors. Fixing gauge and 
global symmetries only $2M$ components of each block survive. Let us consider as an example, the theory with 
$M=1$. We choose
as non vanishing elements $(X_{34})^1_2, (X_{34})^2_3, (Y_{34})^2_1, (Y_{34})^1_2,(Z_{34})^2_2, (Z_{34})^1_3$.  
The anti-fundamental flavors can be represented as a column $\left(X_{13},X_{23},Y_{23}\right)$. The first 
block behaves as $n_f=M$ flavors while the other two as $n_f=4M$ each. In the $M=1$ case, we can take as non vanishing 
components $(X_{13})^1_1,(X_{23})^1_1, (X_{23})^2_{2},(Y_{23})^{3}_{1}, (Y_{23})^{4}_{2}$, the latter being consistent 
with the choices on node $(2)$.
\item Node $(4)$ has $N_f=6M$, $N_c=3M$. The fundamental flavors can be represented by the row 
$\left(X_{42},X_{41},Y_{41}\right)$, 
the first block behaving as $n_f=4M$ flavors, the other two as $n_f=M$. In the case $M=1$ we can take 
$(X_{42})^1_{2},(X_{42})^{2}_{3} ,(X_{42})^{3}_{4}, (X_{41})^1_1,(Y_{41})^3_1$. The anti-fundamental flavors are a column 
$\left(X_{34},Y_{34},Z_{34}\right)$ and a consistent gauge fixing save the same components as appearing for 
node $(3)$ above.
\item Finally node $(1)$ has $N_f=6M$ and $N_c=M$. A consistent gauge fixing on the fundamentals $X_{12}, X_{13}$ and 
the anti-fundamentals $X_{41}, Y_{41}$ selects the same components as the ones just selected before.
\end{itemize}

We are now ready to examine the F-term equations induced from the superpotential (\ref{totalW}). Let us first consider 
those related to $\xi, B, {\tilde B}$. 

One class of solutions of these three equations is $B={\tilde B}=0, \det {\cal M}=\Lambda^{8M}$, i.e. the 
mesonic branch. Let 
us show that this class is in fact empty. The F-term equations w.r.t. $Y_{34}$ and  $X_{34}$ give
\be
M^X_{34}= X_{41}X_{13} , \qquad M^Y_{34}= Y_{41}X_{13}.
\label{f1}
\ee
From the parameterizations introduced previously, the components of $M^X_{34}$
saved by the diagonalization are put to zero by the above equations. Thus $\det {\cal M}=0$, in contradiction 
with $\det {\cal M}=\Lambda^{8M}$.

The only other class of possible solutions is the one with $\xi=\det {\cal M}=0$, $B{\tilde B}=-\Lambda^{8M}$, i.e. 
the baryonic branch. As in the calculation in section \ref{selfsim}, the massive mesons $M^X_{34}, M^Y_{34}$ can 
be integrated out by means of (\ref{f1}). The theory 
flows in the IR, under the scale $\Lambda$ of the node $(2)$, towards a theory 
where node $(4)$ is at strong coupling. The superpotential is
\be
W=\mbox{Tr}\;[N^X_{31}M^Y_{13}-N^Y_{31}M^X_{13}]+ W_{ADS},
\label{finalW}
\ee
where the Affleck-Dine-Seiberg superpotential $W_{ADS}$ involving the mesonic fields 
${\cal N}= \left( N^X_{31}~,~N^Y_{31}\right)\equiv\left(Z_{34} X_{41}~,~Z_{34} Y_{41}\right)$ reads 
\be
W_{ADS}=(N_c-N_f)\left({\Lambda^{3N_c-N_f}\over \det {\cal N}}\right)^{{1\over N_c-N_f}} = 
M\left({\Lambda^{7M}\over \det {\cal N}}\right)^{{1\over M}}.
\ee
It is non-perturbatively generated because of the fact that node $(4)$ has now $N_f(=2M)$ 
less than $N_c (=3M)$. The IR quiver 
triangle is reported in figure 2. 
\begin{figure}[ht]
\begin{center}
{\includegraphics{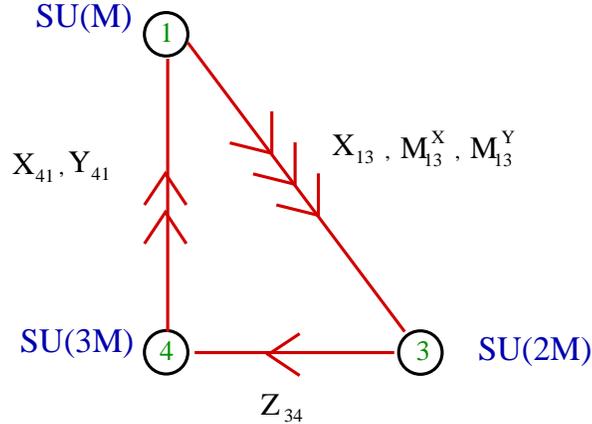}}
\caption{\small The quiver triangle at the bottom of the cascade.}
\label{reg1}
\end{center}
\end{figure}

It is now easy to check that the superpotential (\ref{finalW}) admits no supersymmetric vacuum. 
Indeed, the F-term equations for the $M_{13}$ fields imply $N_{31}=0$, which are not consistent solutions 
for the remaining equations. Hence, as anticipated, supersymmetry is dynamically broken in this model. This also implies  
that the dual geometric deformation that should capture the IR dynamics of the non-conformal $Y^{2,1}$ cascade should be 
non-supersymmetric. This agrees with the statement in \cite{FHU} that there are no complex deformations for the 
complex cone over the first del Pezzo surface.

Our analysis is quite generic and although the corresponding explicit computations might be more and more cumbersome, 
we believe the same conclusion to apply to the full class of  $Y^{p,p-1}$ and $Y^{p,1}$ manifolds, of 
which the case we have considered can be seen as a master example. In fact, in \cite{FHU} it was 
shown that a supersymmetric complex deformation of the cone over $Y^{p,q}$ for any possible value of $q$ 
cannot occur. This suggests that the full $Y^{p,q}$ series might undergo dynamical supersymmetry breaking, although 
the IR dynamics for generic $q$ has not been fully understood, yet.

\vskip 15pt
\centerline{\bf Acknowledgments}
\vskip 10pt
\noindent
We would like to thank Alberto Zaffaroni for useful comments at the beginning of this project. We are also deeply  
in debt with Gabriele Ferretti for a lot of discussions, exchange of ideas and enlightening comments. This work 
is partially  supported by the European Commission RTN Program MRTN-CT-2004-005104, MRTN-CT-2004-503369, 
CYT FPA 2004-04582-C02-01, CIRIT GC 2001SGR-00065
and by MIUR. M.B. is also supported by a MIUR fellowship within the program 
``Incentivazione alla mobilit\`a di studiosi stranieri e italiani residenti all'estero''.


\end{document}